\documentclass[final,5p,times]{elsarticle}
\pdfoutput=1
\usepackage{amssymb}
\usepackage{graphicx}
\usepackage{dcolumn}
\usepackage{bm}
\usepackage[mathlines]{lineno}
\usepackage{amsmath}
 \usepackage{latexsym}
\usepackage{url} 
\usepackage{color}
\journal{Physics Letters B}
\begin{document}
\begin{frontmatter}
\title{Neutron spectroscopy of $^{26}$Mg states:  constraining the stellar neutron source $^{22}$Ne($\alpha, n$)$^{25}$Mg}
\author[1,2]{C.~Massimi}
\author[3]{S.~Altstadt}
\author[4]{J.~Andrzejewski}
\author[5]{L.~Audouin}
\author[6]{M.~Barbagallo}
\author[7]{V.~B\'{e}cares}
\author[8]{F.~Be\v{c}v\'{a}\v{r}}
\author[9]{F.~Belloni}
\author[9]{E.~Berthoumieux}
\author[10]{J.~Billowes}
% ...
\address[1]{Istituto Nazionale di Fisica Nucleare, Bologna, Italy}
\address[2]{Dipartimento di Fisica e Astronomia, Universit\`a di Bologna, Italy}
\address[3]{Johann-Wolfgang-Goethe Universit\"{a}t, Frankfurt, Germany}
\address[4]{Uniwersytet \L\'{o}dzki, Lodz, Poland}
\address[5]{Institut de Physique Nucl\'eaire, CNRS-IN2P3, Univ. Paris-Sud et Paris-Saclay, 91406 Orsay Cedex, France}
\address[6]{Istituto Nazionale di Fisica Nucleare, Bari, Italy}
\address[7]{Centro de Investigaciones Energeticas Medioambientales y Tecnol\'{o}gicas (CIEMAT), Madrid, Spain}
\address[8]{Charles University, Prague, Czech Republic}
\address[9]{Commissariat \`{a} l'\'{E}nergie Atomique (CEA) Saclay - Irfu, Gif-sur-Yvette, France}
\address[10]{University of Manchester, Oxford Road, Manchester, UK}
% ...
%
\begin{abstract}
This work reports on accurate, high-resolution measurements of the $^{25}$Mg($n, \gamma$)$^{26}$Mg and $^{25}$Mg($n, tot$) 
cross sections in the neutron energy range from thermal to about 300 keV, leading to a significantly improved 
$^{25}$Mg($n, \gamma$)$^{26}$Mg parametrization. The relevant resonances for $n+^{25}$Mg  were characterized from a combined 
R-matrix analysis of the experimental data. This resulted in an unambiguous  spin/parity assignment of the corresponding 
excited states in $^{26}$Mg. With this information experimental upper limits of the
 reaction rates for $^{22}$Ne($\alpha, n$)$^{25}$Mg and $^{22}$Ne($\alpha, \gamma$)$^{26}$Mg were established, potentially leading to a significantly higher ($\alpha, n$)/($\alpha, \gamma$) ratio than previously evaluated. The impact of these results have been studied for stellar models in the mass range 2 to 25 $M_{\odot}$.
\end{abstract}
\begin{keyword}
s process \sep $\alpha+^{22}$Ne \sep  neutron spectroscopy 
\PACS 26.20.Kn \sep 25.40.Ny
\end{keyword}
\end{frontmatter}
\section{Introduction} \label{Introduction}
The stellar nucleosynthesis of elements heavier than iron occurs
via neutron capture reactions and subsequent $\beta$ decays. The so-called
slow neutron capture process or $s$ process~\cite{Kaep} takes place
in low and intermediate mass stars during their Asymptotic Giant
Branch (AGB) phase. 
An additional $s$-process contribution comes in
the helium-burning core and
subsequent carbon-burning shell phases of massive stars.
The $^{22}$Ne($\alpha, n$)$^{25}$Mg reaction constitutes a major neutron source 
for the $s$ process. In AGB stars, this source is activated during a
series of convective He-shell burning episodes, so-called thermal pulses,
with temperatures as high as 400 MK (more than 30 keV thermal energy).
Although it contributes only about 5\% to the total neutron budget, this
scenario determines the final abundance pattern. Hence, it is 
decisive for deriving information on neutron density, temperature, and 
pressure in the He-burning layers of AGB stars~\cite{Kaep}.
In massive stars, where thermal energies of 25 and 90 keV are reached 
during the He-core and C-shell burning phases, the neutron budget is by
far dominated by the $^{22}$Ne($\alpha, n$)$^{25}$Mg reaction.

The $^{22}$Ne($\alpha, n$)$^{25}$Mg cross section in the temperature regime of 
the $s$ process is affected by unknown resonance contributions. 
Also the competing $^{22}$Ne($\alpha, \gamma$)$^{26}$Mg reaction, whose cross section is 
also insufficiently known, adds to the persistent uncertainty of the neutron 
budget and the associated neutron densities during stellar nucleosynthesis. 
The ($\alpha, \gamma$) reaction contributes to the destruction of $^{22}$Ne 
during the entire helium burning phase because of its positive $Q$-value 
of 10.615 MeV, already before the threshold for neutron production via 
$^{22}$Ne($\alpha, n$)$^{25}$Mg ($Q=-478$ keV) is reached. In this context, the
$^{25}$Mg($n, \gamma$)$^{26}$Mg reaction plays a non-negligible role as well, 
because the high $s$ abundance of $^{25}$Mg makes it a significant 
neutron poison for the  $s$ process. So far, this reaction rate is too uncertain
for a quantitative assessment of the poisoning effect ~\cite{Mass}.

Experimental data for the $^{22}$Ne($\alpha, n$)$^{25}$Mg cross section are essentially 
limited to $\alpha$ energies above about 800 keV. At lower energies the most 
recent direct measurements ~\cite{Drot,Jaeg,Giesen} provided only an upper 
limit of about $10^{-11}$ barn, corresponding to the experimental sensitivity. 
The lowest resonance observed in direct measurements is at $E_{\alpha}^{Lab}=832\pm2$ 
keV ({\it e. g.} Ref.~\cite{Jaeg}).   
\section{Indirect approach}\label{Indirect}
Since a direct approach is exceedingly difficult, indirect and transfer reactions 
were considered as an alternative way for constraining the reaction rate at low
energies. For instance, the $\alpha$-transfer reaction $^{22}$Ne($^6$Li, d)$^{26}$Mg 
has been studied \cite{Giesen,Ugalde,Talwar} to search for levels  in 
$^{26}$Mg also below the neutron threshold at 11.093 MeV. These studies are 
particularly relevant for estimating the $^{22}$Ne($\alpha, \gamma$)$^{26}$Mg 
rate, where levels below and above the neutron threshold are involved. Direct $^{22}$Ne($\alpha, \gamma$)$^{26}$Mg measurements have only been carried out above 0.8 MeV. In view of the scarce experimental information,
evaluations had, therefore, to rely on theoretical estimates. Accordingly,
the uncertainties of the stellar $^{22}$Ne($\alpha, n$)$^{25}$Mg and $^{22}$Ne($\alpha, 
\gamma$) rates are still dominated by unknown properties of states in the 
compound nucleus $^{26}$Mg above the $\alpha$ threshold at 10.615 MeV. 
\begin{figure}[h]
\includegraphics[width=0.95\linewidth]{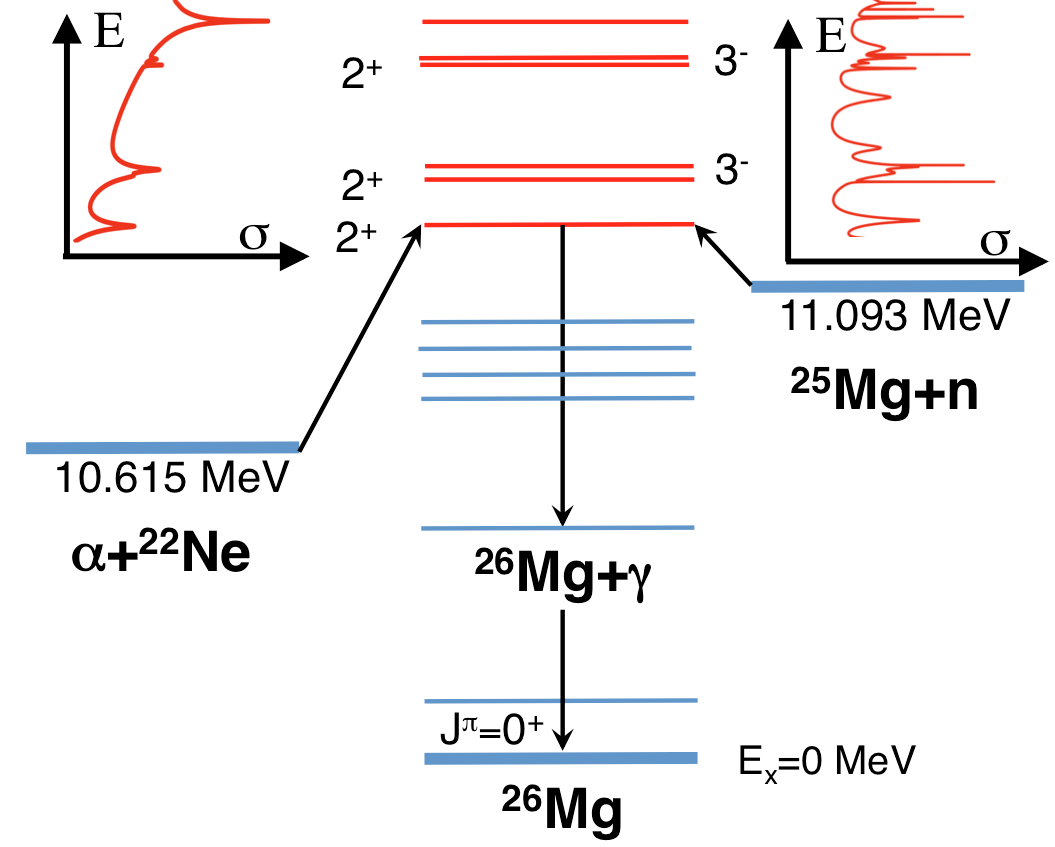}
\caption{\label{fig_levels} Color online. Level scheme (not to scale) of
$^{26}$Mg with the entrance channels $^{22}$Ne+$\alpha$ and 
$n+^{25}$Mg and the exit channel $^{26}$Mg+$\gamma$. Excited states 
in $^{26}$Mg with natural parity ($0^+$, $1^-$, $2^+$, ...) and energies below $E_{\alpha}^{Lab}=
850$ keV (in red) can be studied via neutron resonance spectroscopy
of $n+^{25}$Mg as indicated by the schematic $^{25}$Mg($n, \gamma$)$^{26}$Mg
cross section.}
\end{figure}

As indicated in Fig.~\ref{fig_levels}, these states can be studied 
via neutron resonance spectroscopy of $n+^{25}$Mg. In particular, a 
simultaneous resonance shape analysis of total and capture cross section 
data can be used to determine the resonance energies, $E_R$, spin
parities, $J^\pi$, as well as the capture and neutron widths $\Gamma_\gamma$ 
and $\Gamma_n$~\cite{Frohner}. Information from previous $^{25}$Mg($n, \gamma$)$^{26}$Mg 
measurements suffers from considerable uncertainties due to sample 
problems and counting statistics. Moreover,  the available transmission 
data have been obtained with natural Mg samples, a serious limitation 
in view of the dominant abundance of $^{24}$Mg in natural
magnesium. The corresponding differences between present and previous
transmission results are discussed in Ref.~\cite{EPJproc}.
\section{Measurements of $n+^{25}$Mg}\label{Measurements}
 This work on $n+^{25}$Mg 
 combined two neutron time-of-flight measurements, using the 
CERN-n\_TOF facility~\cite{nTOF} for the ($n, \gamma$) channel
and the GELINA accelerator at IRMM/Geel ~\cite{GELINA} for the 
total cross section. To avoid the limitations of previous attempts, both measurements were carried out with highly 
enriched, metallic samples (97.87\% $^{25}$Mg). 
The n\_TOF facility, which provides a
white neutron spectrum from thermal energy to about 1-GeV neutron
energy, was chosen because it is particularly well suited 
for accurate high-resolution neutron measurements due to its outstanding 
instantaneous neutron flux (about $10^6$ neutrons per bunch at the experimental area) and the very long flight-path of 185 m.
The experimental setup was based on two deuterated benzene
(C$_{6}$D$_{6}$) scintillation detectors (cylindrical cells with an active volume of about 1 liter), placed on either side of the 
neutron beam, 9.0 cm upstream of the sample. These detectors are characterized by a very small neutron sensitivity.
The chosen geometrical configuration minimized
the background due to scattered in-beam $\gamma$ rays and reduced systematic effects due to the anisotropy in the 
angular distribution from primary $\gamma$-rays
following $p-$wave neutron  resonances. The total number of neutrons impinging on the various samples used in the measurement was determined by means of a low-mass, $^{6}$Li-based, neutron detector~\cite{SiMon}, while the energy dependence of the neutron flux derived from a dedicated study~\cite{Barb}.
The well-established total energy detection principle in combination with the pulse height weighting technique~\cite{NDS}  was applied
to make the detection efficiency for a capture event directly proportional to the total $\gamma$-ray energy available in
the capture event. Moreover the saturated resonance technique~\cite{NDS} was used to normalize the capture data.
The measurement covered the energy range from thermal energy to about 300 keV with a resolution 
of 0.05\% to 0.3\% in the relevant energy range between 1 and 300 keV. 

The total cross section of $n+^{25}$Mg was measured with a $^6$Li-glass
detector at the 50 m station of the GELINA facility for energies up to about 300 keV.
A combination of Li-carbonate plus resin, Pb and Cu-collimators was used to reduce the neutron beam to a diameter of less than 35 mm at the sample position.
 The sample was placed in an automatic sample changer at a distance of 23.78 m from the neutron source. Close to the
sample position a $^{10}$B anti-overlap filter was placed to absorb slow neutrons from a
previous burst.  In order to have a well-characterized measurement condition, permanent Na and Co black resonance filters~\cite{NDS} were used to continuously monitor the background at 2.85 keV and 132 eV. Additional black resonance filters were used in dedicated runs. Thanks to the small dimensions of the neutron-producing target and the 
excellent time characteristic (1-ns pulse width), GELINA is particularly suited for high-resolution 
transmission measurements in the keV region.   

Essential improvements with respect to the previous capture measurement 
at n\_TOF~\cite{Mass} are the quality of the highly enriched $^{25}$Mg 
sample and a substantially reduced background of in-beam $\gamma$ 
rays~\cite{nTOF}. The total cross section measurement benefits also 
significantly from the highly enriched sample. In both measurements the 
experimental background was determined in dedicated runs.

\begin{figure}[h]
\includegraphics[width=0.95\linewidth]{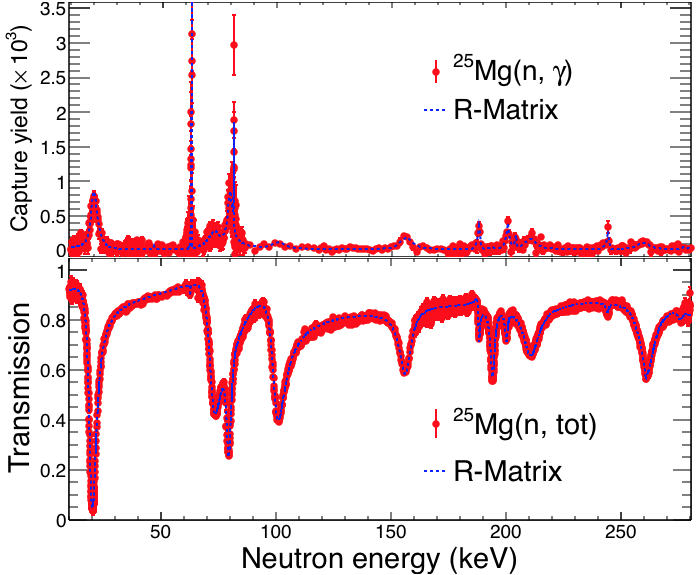}
\caption{\label{fig_captra}Color online. Measured neutron capture yield and transmission 
of $^{25}$Mg. Experimental data are represented by red symbols and 
 the simultaneous R-matrix of both data sets is given by dashed lines.}
\end{figure}

The data reduction for the determination of the cross section starting from detectors counting rate was performed following the procedures described in~\cite{NDS}.
The capture yield (representing the fraction of the neutron beam undergoing capture events) obtained at n\_TOF and the transmission (representing the fraction of neutron beam that traverses the sample without interaction) from GELINA are presented in Fig.~\ref{fig_captra} and compared with the results of a simultaneous R-matrix 
 analysis of both data sets. More in detail, the experimental capture yield $Y$, deduced from the response of the capture detection system and the flux-monitor is related to the total and capture cross-sections, $\sigma_{tot}$ and $\sigma_\gamma$ respectively, by the following expression~\cite{NDS}: 
\begin{equation}
Y(E_n)=(1-e^{-n_C\sigma_{tot}(E_n)})\frac{\sigma_\gamma(E_n)}{\sigma_{tot}(E_n)} + Y_{MS}(E_n),
\label{eq:CapYield}
\end{equation} 
while the transmission  $T$, which was experimentally obtained from the ratio of Li-glass spectra resulting from a sample-in and a sample-out measurement, is related to the total cross-section by~\cite{NDS}: 
\begin{equation}
T(E_n)=e^{-n_T\sigma_{tot}(E_n)},
\label{eq:TRA}
\end{equation}
where $n_C=(3.012\pm0.009)\times10^{-2}$ atoms/b denotes the areal density of the capture sample and 
 $n_T=(5.02\pm0.04)\times10^{-2}$ atoms/b the one of the transmission sample, $E_n$ the laboratory neutron energy, and $Y_{MS}$ the contribution of multiple interaction in the sample. This latter term, together with other experimental effects such as self-shielding, Doppler broadening and response of the time-of-flight spectrometer, are properly taken into account  in the 
R-Matrix codes SAMMY~\cite{SAMMY} and REFIT~\cite{REFIT} used in the present combined analysis.

\section{Improved $^{25}$Mg($n, \gamma$)$^{26}$Mg cross section}
Resonance energies, $E_R$, partial widths, $\Gamma_\gamma$ 
and $\Gamma_n$, and the spins and parities, $J^{\pi}$ were determined 
by a combined R-matrix analysis in the energy region $E_n < 300$ keV, corresponding to an
incident $\alpha$-particle energy from threshold at $E_\alpha =570$
keV to approximately 850 keV, where direct ($\alpha,n$) measurements are difficult and 
accurate cross section data are missing. 

The information on $J^{\pi}$ is particularly important for identifying the 
states in $^{26}$Mg that contribute to the $^{22}$Ne($\alpha, n$)$^{25}$Mg 
cross section. This selection is crucial because $^{22}$Ne nuclei and 
$\alpha$ particles have both $J^\pi=0^+$. Accordingly, ($\alpha, n$) 
reactions can only populate natural-parity states ($0^+$, $1^-$, $2^+$, ...) in $^{26}$Mg, whereas 
the  $n+^{25}$Mg reaction proceeds also via $J^\pi=1^+, 2^-, 3^+, ...$ 
states. 

The results of the combined R-matrix analysis of both cross sections are summarized in Table~\ref{tab_res}, including the deduced firm $J^{\pi}$ assignments. 
For the natural-parity states, the $\alpha$ energies in the laboratory system
$E_{\alpha}^{Lab}$ are given in column 3 for comparison with direct
$\alpha + ^{22}$Ne measurements. The uncertainties on $E_x$ and
$E_{\alpha}^{Lab}$ are determined by the uncertainty in neutron
energy, which never exceeds 0.1 keV. This is an important result in itself 
because it allows one to resolve inconsistencies in charged-particle data. 

\begin{figure}[h]
\includegraphics[width=0.95\linewidth]{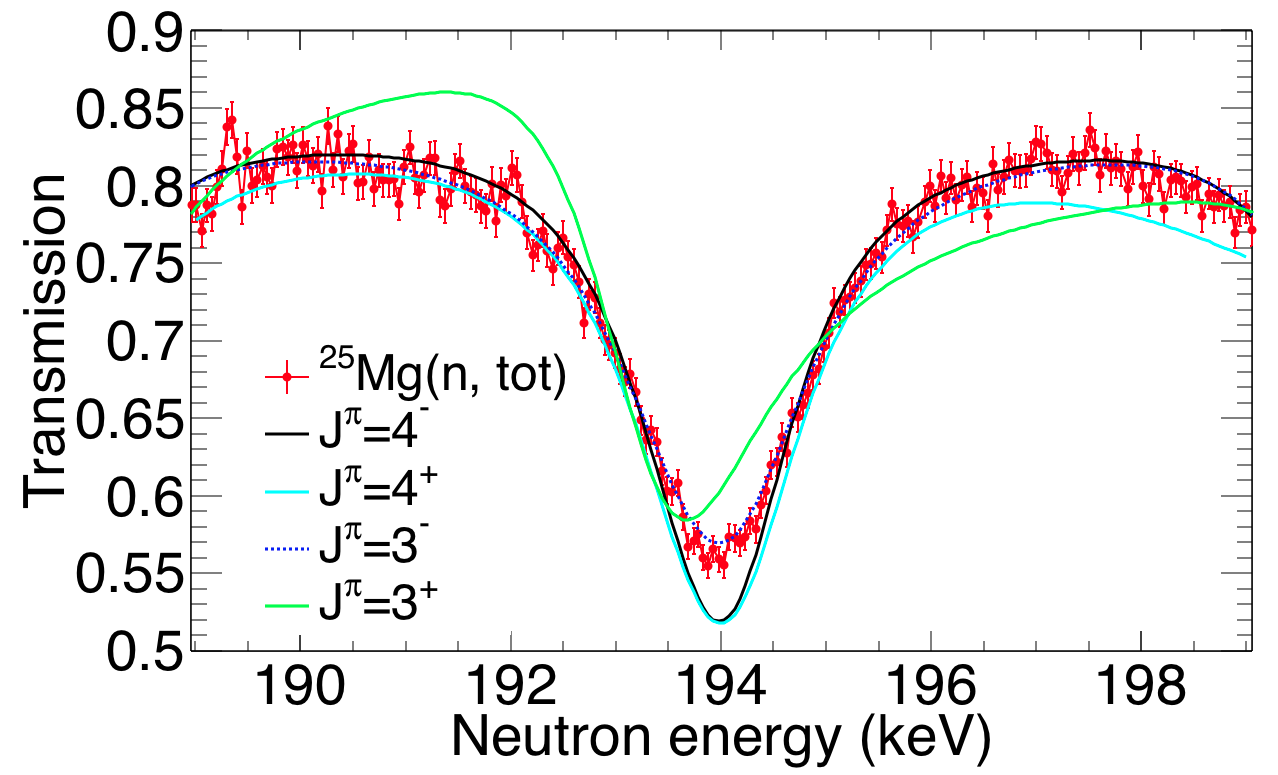}
\caption{\label{fig_adj} Color online. Resonance shape analysis of the neutron resonance at $E_n=194$ keV. The spin parity assignment $J^{\pi}=4^-$~\cite{Koeh}  is excluded by the present data, which clearly indicate $J^{\pi}=3^-$. Only uncorrelated uncertainties, attributable to counting statistics, are reported. }
\end{figure}

Fig.~\ref{fig_adj} demonstrates the sensitivity of the present data to spin and parity, which clearly exclude the previous $J^{\pi}=4^-$ assignment~\cite{Koeh} for the neutron resonance at 194 keV. In particular, the lowest values of the chi-squared per number of degrees of freedom of the fit, were 1.1, 5.6, 7.3 and 17 assuming $J^{\pi}=3^-$, 4$^+$, 4$^-$ and 3$^+$ respectively.  
In this manner the $J^{\pi}$ values were determined for each resonance. In two cases the spin-parity assignment was changed with respect to the previous evaluation. As shown in Ref.~\cite{EPJproc}, these two resonances could not be accurately studied by using the transmission data on natural magnesium ({\it e. g.} reported in~\cite{Koeh}) since the experimental signature was dominated by strong resonances in $n+^{24}$Mg.

Compared to evaluated resonance data~\cite{Mugh06}, six neutron 
resonances had to be removed in the energy range covered in 
Table~\ref{tab_res}. These weak resonances ($\Gamma_n \approx 0.5$ eV) were assigned in previous 
$n+^{25}$Mg experiments, but were not identified in this work. 

\begin{table}
\caption{$n+^{25}$Mg resonance parameters and corresponding
excitation energies of the $^{26}$Mg compound nucleus. The  quoted  uncertainties  were
obtained by the R-Matrix fit.
\label{tab_res}}
 \begin{tabular}{lccccc}
 \hline
$E_n$ & $E_x$ & $E_{\alpha}^{Lab}$ &  $J^{\pi}$ & $\Gamma_\gamma$ & $\Gamma_n$     \\
(keV) & (keV) & (keV) & ($\hbar$)  & (eV) & (eV)     \\
\hline
$19.92(1)$ &11112 &589  & $2^+$&$1.37(6)$ & $2095(5)$ \\
$62.73(1)$ &11154 & &  $1^+$&$4.4(5)$ & $7(2)$ \\
$72.82(1)$ &11163 &649  & $2^+$&$2.8(2)$ & $5310(50)$ \\
$79.23(1)$ &11169 &656  & $3^-{}^{(a)}$&$3.3(2)$ & $1940(20)$ \\
$81.11(1)$ &11171 & &   &$5(1)$ & $1-30$ \\
$100.33(2)$ &11190 &  & $3^+$ & $1.3(2)$ & $5230(30)$ \\
$155.83(2)$ &11243 &  & $2^-$ & $4.7(5)$ & $5950(50)$ \\
$187.95(2)$ &11274 & 779 & $2^+$ & $2.2(2)$ & $410(10)$ \\
$194.01(2)$ &11280 &786  & $3^-{}^{(a)}$ & $0.3(1)$ & $1810(20)$ \\
$199.84(2)$ &11285 & & $2^-$ & $4.8(4)$ & $1030(30)$ \\
$203.88(4)$ &11289 & &  & $0.9(3)$ & $3-20$ \\
$210.23(3)$ &11295 &  & $2^-$ & $6.6(6)$ & $7370(60)$\\
$243.98(2)$ &11328 & (843) &  & $2.2(3)$ & $171(6)$\\
$260.84(8)$ &11344 & & &$1.0(2)$ & $300-3900$\\
$261.20(2)$ &11344 & & $>3$ &$3.0(3)$ & $6000-9000$\\
 \hline
 \end{tabular}
\begin{center}
$^{(a)}$ Parity change with respect to previous evaluations.\\
  \end{center}\end{table}
Below the lowest directly observed resonance in the  $^{22}$Ne($\alpha,
n$)$^{25}$Mg reaction cross-section at $E_{\alpha}^{Lab}\approx 830$ keV 
($E_n\approx235$ keV), five natural-parity states have been identified 
corresponding to energies $E_{\alpha}^{Lab}=589$, 649, 656, 779 and 
786 keV. The doublets at 649/656 and 779/786 keV were resolved 
owing to the good energy resolution in the neutron channel, which is
significantly better than in charged-particle experiments. 
The lowest observed $^{22}$Ne($\alpha, n$)$^{25}$Mg resonance at $E_{\alpha}^{Lab}=832\pm2$ keV~\cite{Jaeg} was not observed in the present neutron data at $E_n=234$ keV, while a resonance with a width compatible to the one reported in~\cite{Jaeg} ($\Gamma=250\pm170$ eV) is located at $E_n=243.98$ keV, {\it i.e.} $E_{\alpha}^{Lab}= 843$ keV. One possible explanation is that the two resonances correspond to the same state, in which case the energies are inconsistent with each other. Another possibility is that they correspond to two different levels: in this case the width of the $(\alpha, n)$ resonance at 832 keV is lower than quoted in~\cite{Jaeg}, being below the sensitivity of the present measurements ($\Gamma \approx 20$ eV), while the observed neutron level is not a natural parity state.

The $^{25}$Mg($n, \gamma$)$^{26}$Mg cross section was also significantly 
improved by an R-matrix analysis of the combined data sets for $^{25}$Mg($n, 
\gamma$) and ($n, tot$), resulting in rather accurate Maxwellian averaged 
cross section (MACS) for thermal energies between $kT=5$ and 100 keV. 
Though higher resonances up to 900 keV were considered as well, the MACS values are dominated by contributions from the 
resonances in Table~\ref{tab_res}. The present results and the currently 
recommended values in the KADoNiS~\cite{KADONIS} compilation and in 
Ref.~\cite{Mass} are compared in Table~\ref{tab_MACS}. While the new
MACS values are compatible with the previous measurement~\cite{Mass} except for the lowest and highest temperature, the 
uncertainties were reduced by almost a factor of three on average. These uncertainties are the sum of uncorrelated
or statistical uncertainties and systematic uncertainties. 
\begin{table*}[hbtp]
\begin{center}
\caption{$^{25}$Mg(n, $\gamma$) Maxwellian-averaged cross sections
(in mb), compared with a previous work and recommended values. Experimental values include the contributions from direct radiative capture \cite{Mass}.
 \label{tab_MACS}}
  \begin{tabular}{lccccccccccc}
\hline
$kT$ (keV) &5&10&15&20&25&30&40&50&60&80&100 \\
\hline
KADoNiS &4.8&5.0&5.5&6.0&6.2&$6.4(4)$&6.2&5.7&5.3&4.4&3.6\\
Ref.~\cite{Mass} &$3.5(4)$&$5.1(6)$&$4.9(6)$&$4.6(4)$&$4.4(6)$&$4.1(6)$&$3.5(6)$&$2.9(5)$&$2.5(4)$&$1.9(3)$&$1.4(2)$\\
this work  &$2.8(2)$&$4.4(2)$&$4.3(2)$&$4.2(2)$&$4.0(2)$&$3.9(2)$&$3.6(2)$&$3.4(2)$&$3.0(2)$&$2.5(3)$&$2.2(3)$\\
 \hline
 \end{tabular}
\end{center}
 \end{table*}
\section{Impact on reaction rate calculations}\label{Constraints}
As mentioned above, the reaction channels for $\alpha+^{22}$Ne and 
$n+^{25}$Mg open at excitation energies $E_x=10.615$ and 11.093 MeV 
in the compound nucleus $^{26}$Mg. Apart from the levels between 
$\alpha$- and $n$-channel, the present study provides access to the 
relevant states in the energy region between the $^{22}$Ne($\alpha, n$)$^{25}$Mg 
threshold and the lowest resonance reached in direct $\alpha+^{22}$Ne 
experiments. Therefore the new resonance information has been used for calculating the upper limit of the reaction rates, based purely on experimentally-available information for $^{22}$Ne($\alpha, n$)$^{25}$Mg and $^{22}$Ne($\alpha, \gamma$)$^{26}$Mg. In particular,
they were defined using: (i) data from an $\alpha$-transfer 
reaction for the resonance below the neutron threshold~\cite{Talwar}, (ii) data 
from this work, and (iii) data from direct measurements above 
$E_{\alpha}^{Lab}=800$ keV \cite{Jaeg}. The calculation, based on the simple narrow-resonance formalism~\cite{CC}, 
requires as input 
the resonance energies and the so-called kernels (also referred to as resonance strength), $k_n$ or $k_\gamma$, 
\begin{equation}
\begin{split}
k_n & =g\frac{\Gamma_\alpha \Gamma_n}{\Gamma}, \\
 k_\gamma & =g\frac{\Gamma_\alpha \Gamma_\gamma}{\Gamma}=k_n\frac{\Gamma_\gamma}{\Gamma_n};
\label{eq:Res_stre}
\end{split}
\end{equation}
where $\Gamma$ is the total width and $g=(2J+1)/[(2I+1)(2i+1)]$ the  
statistical factor and  $J$, $I$ and $i$ the spin of the resonance, target and projectile, respectively.
Because $\Gamma_\alpha\ll\Gamma_\gamma\ll\Gamma_n$, 
the kernels can be reduced to $k_n=g\Gamma_\alpha$. 

Apart from $\Gamma_\alpha$, all quantities were determined in this work.
For the present analysis we estimated the upper limits  of $\Gamma_\alpha$ based on the 
experimental constraint on the  $^{22}$Ne($\alpha, n$)$^{25}$Mg cross section ($\le 10^{-11}$ barn, corresponding to $\Gamma_\alpha \approx 10^{-8}$ eV).

The contributions of the five resonances below $E_{\alpha}^{Lab}\approx 
800$ keV to the $^{22}$Ne($\alpha, n$)$^{25}$Mg rate depend on the temperature of the specific s-process site.
Particularly important is the 30\% decrease of the upper limit of the rate compared to Ref.~\cite{Talwar} around $kT=25$ keV (0.25 to 0.3 GK in Fig. 4), which affects the neutron economy  during the He shell burning in low-mass AGB stars and during core He burning in massive stars. 
  
\begin{figure}[h]
\includegraphics[width=0.95\linewidth]{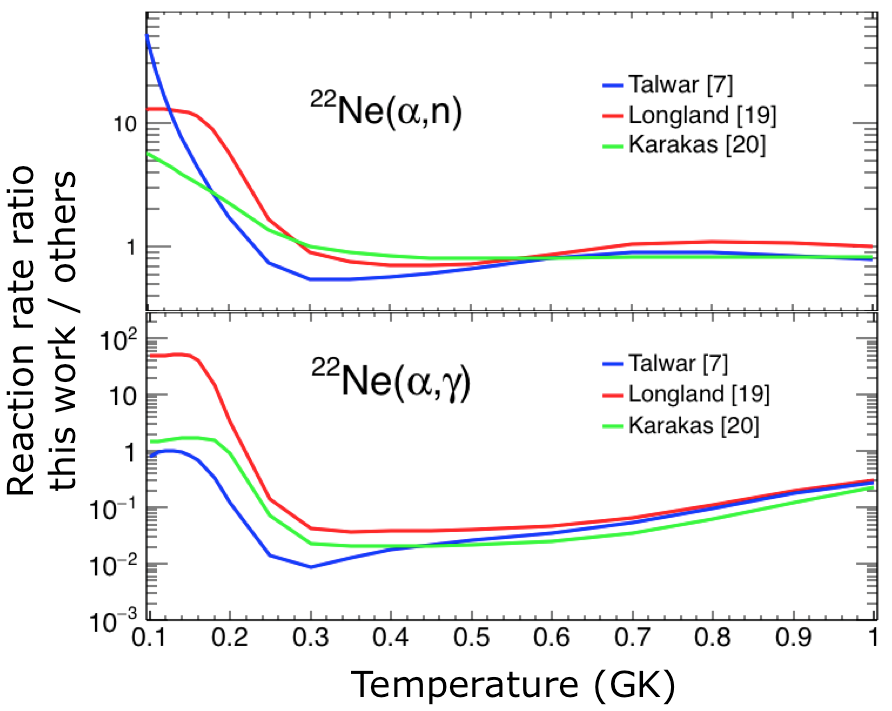}
\caption{\label{fig_Neag}Color online. 
Reaction rate ratio of the upper limits for the $^{22}$Ne($\alpha, n$)$^{25}$Mg (top panel) from this work compared to the upper limits in Refs.~\cite{Talwar,Karakas,LONG}. The same ratios are presented for the $^{22}$Ne($\alpha, \gamma$)$^{26}$Mg (bottom panel). Below 0.3 GK the present data are indicating an enhanced ($\alpha, n$) rate, whereas the ($\alpha, \gamma$) rate is strongly reduced above that temperature.}
\end{figure}

The neutron resonance data obtained in this work were also used to determine the upper limit of the $^{22}$Ne($\alpha, \gamma$)$^{26}$Mg rate. As shown in Fig.~\ref{fig_Neag},  these results differ significantly compared to recent evaluations~\cite{Talwar,LONG,Karakas}, also in the important temperature range above 0.3 GK. Part of these discrepancies were caused by a misinterpretation of previous neutron data, in particular for the doublet at $E_n=79$ keV.  While the impact of the 79-keV doublet on the $^{22}$Ne($\alpha, n$)$^{25}$Mg  rate is relatively small (at most 3\%), it plays a major role for the ($\alpha, \gamma$) rate. As discussed in Ref.~\cite{Talwar}, $\alpha$-transfer measurements provide experimental evidence of a natural parity state in $^{26}$Mg corresponding to an energy of about $E_n=80$ keV. Prior to this measurement the resonance at E$_n=79$ keV was tentatively assigned to be of unnatural parity whereas the closest resonance at $E_n=81$ keV was considered as the one with natural parity. This misinterpretation resulted in a biased $^{22}$Ne($\alpha, \gamma$)$^{26}$Mg reaction rate, since the resonance strength of the ($\alpha,\gamma$) channel scales with the $\Gamma_n/\Gamma_\gamma$ ratio (see Eq.~\ref{eq:Res_stre}). With the present spin-parity assignment, the contribution to the reaction rate of this level became negligible since the $\Gamma_n/\Gamma_\gamma$ ratio of the resonance at E$_n=79$ keV is about a factor of 2200 larger than the ratio of the resonance at E$_n=81$ keV.

 \begin{table}[hbt]
\caption{Upper limits of the reaction rates, given in cm$^3$/(mole$\cdot$s), determined from the present measurements. \label{tab_RATES}}
\begin{center}
 \begin{tabular}{ccc}
\hline
Temperature  &$^{22}$Ne($\alpha,n$)&$^{22}$Ne($\alpha,\gamma$)\\
 10$^9$ K & \multicolumn{2}{c}{upper limits} \\
 \hline
0.06 & 2.97$\times10^{-44}$ & 4.13$\times10^{-21}$\\
0.08 & 5.56$\times10^{-34}$ & 8.45$\times10^{-28}$\\
0.10 & 7.56$\times10^{-28}$ &  3.23$\times10^{-24}$\\
0.12 & 8.96$\times10^{-24}$ & 1.55$\times10^{-21}$\\
0.13 &  3.29$\times10^{-22}$& 1.67$\times10^{-20}$\\
0.14 & 7.16$\times10^{-21}$ & 1.27$\times10^{-19}$\\
0.15 & 1.04$\times10^{-19}$ & 7.35$\times10^{-19}$\\
0.16 & 1.07$\times10^{-18}$ & 3.39$\times10^{-18}$\\
0.18 & 5.32$\times10^{-17}$ & 4.27$\times10^{-17}$\\
0.19 & 2.78$\times10^{-16}$ & 1.23$\times10^{-16}$\\
0.20 & 1.25$\times10^{-15}$ & 3.21$\times10^{-16}$\\
0.25 &  4.99$\times10^{-13}$& 1.43$\times10^{-14}$\\
0.30 & 4.52$\times10^{-11}$ & 5.82$\times10^{-13}$\\
0.35 & 1.44$\times10^{-09}$ & 1.82$\times10^{-11}$\\
0.40 & 2.09$\times10^{-08}$ & 2.66$\times10^{-10}$\\
0.45 & 1.74$\times10^{-07}$ & 2.12$\times10^{-09}$\\
0.50 & 1.03$\times10^{-06}$ & 1.15$\times10^{-08}$\\
0.60 &  2.24$\times10^{-05}$ & 1.50$\times10^{-07}$\\
0.70 &  3.36$\times10^{-04}$ & 1.18$\times10^{-06}$\\
0.80 &  3.19$\times10^{-03}$ & 7.38$\times10^{-06}$\\
0.90 & 1.95$\times10^{-02}$ & 3.67$\times10^{-05}$\\
1.00 &  8.45$\times10^{-02}$ & 1.43$\times10^{-04}$\\
 \hline
 \end{tabular}
\end{center}
 \end{table}

In summary, the present resonance analysis of the neutron-induced
cross sections of $^{25}$Mg has led to decisive improvements
for the poorly known states in $^{26}$Mg with excitation energies 
between 11112 and 11344 keV. This concerns accurate level energies, 
firm spin/parity assignments, and the determination of reliable 
$\Gamma_\gamma$ and $\Gamma_n$ values. The conclusive 
identification of their properties were used to determine the
contributions of levels below the energy range covered by direct
($\alpha, n$) measurements. While waiting for the crucial estimate of $\Gamma_\alpha$, which will require a large experimental effort, the present results can lead to improvements in the calculation of the stellar 
$^{22}$Ne+$\alpha$ rates in the astrophysically relevant energy range
substantially. In particular, it was found that the ($\alpha, \gamma$) channel had been strongly overestimated so far. 
The present results are summarized in Table ~\ref{tab_RATES}.
\section{Astrophysical implications and conclusions}
The astrophysical consequences of the present results 
have been studied using the upper limits established in this work. 
Their impact on the weak-s process in massive stars was explored for a 25 $M_{\odot}$~\cite{HIR} star.
The calculated abundances are consistent with the latest evaluations~\cite{Talwar, LONG}. The present
 uncertainty of the $\alpha+^{22}$Ne rate causes large differences in the weak s-process abundances, 
 up to a factor 50 in the Sr region. 

Noticeable changes are also found in intermediate-mass AGB models (IMS-AGBs, 
3$<$M/M$_\odot$$<$7). As in these stars higher temperatures are reached 
during thermal pulses than in low-mass AGB stars, the strong
preponderance of the $^{13}$C($\alpha, n$)$^{16}$O neutron source 
is progressively substituted by the $^{22}$Ne($\alpha, n$)$^{25}$Mg reaction. This is 
particularly evident at low metallicities.

Using the present upper limits for the $^{22}$Ne($\alpha, 
n$) rates, the $s$-process yields of AGB stars between 2 and 5 M$_\odot$ and
an initial iron fraction corresponding to 0.7\% of the solar value were calculated 
for comparison with reference AGB models from the FRUITY data base \cite{cristallo}. 
The resulting $s$ abundances of Y and La were selected for comparison with Pb, 
because Y and La are characterizing the first and second $s$-process abundance 
peak.
\begin{figure}[htb]
\includegraphics[width=0.95\linewidth]{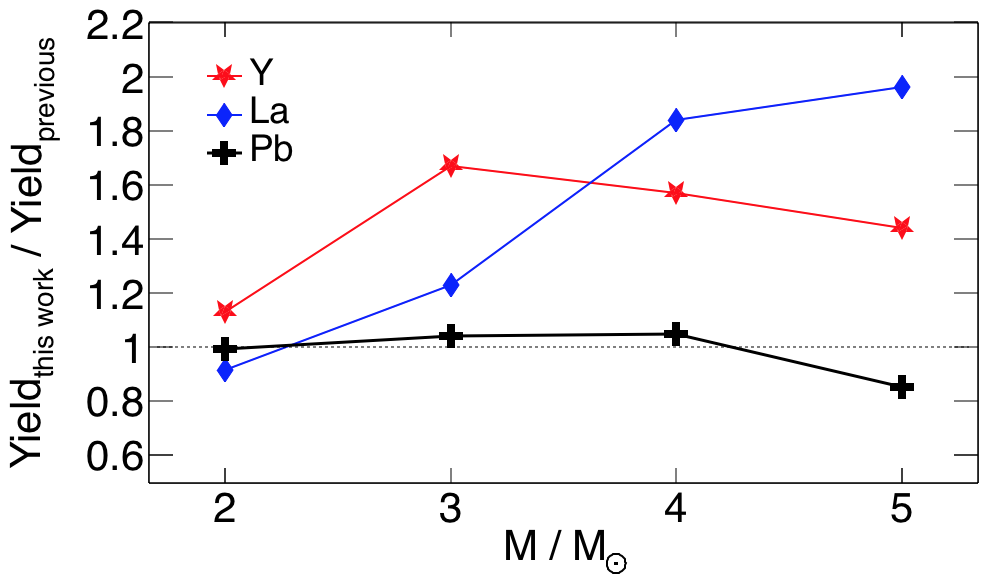}
\caption{\label{fig_astro} Ratios of the $s$ abundance of Y, La, and Pb from stellar 
$s$-process models using the present upper limits of the $^{22}$Ne($\alpha, n$)$^{25}$Mg 
rate and from the FRUITY reference set
\cite{cristallo}.}
\end{figure} 
As illustrated in Fig. \ref{fig_astro} the production of Y and - especially - of La 
is increasing with AGB mass as a consequence of the more efficient 
$^{22}$Ne($\alpha, n$)$^{25}$Mg source, whereas the Pb abundance remains basically 
frozen. For a fixed amount of Pb we obtain, therefore, larger abundances in the 
first and, especially, in the second $s$-process peak. This pattern in higher-mass 
stars may help to explain the surface distributions in $s$-process enriched globular 
cluster stars, which exhibit a comparably low Pb abundance \cite{straniero}. Whether 
a mix of enhanced $s$ contributions from  IMS-AGBs at the expense of the Pb 
producing LMS-AGBs could provide a plausible solution will be analyzed in a 
dedicated paper. 
\section*{Acknowledgments}
The isotope used in this research were supplied by the United States Department of Energy Office of Science by the Isotope Program in the Office of Nuclear Physics. 

M.~Pignatari and I.~Van~Rijs thank NuGrid for the support.

\end{document}